\documentclass{aa}

\usepackage{natbib}

\bibpunct{(}{)}{;}{a}{}{,} 

\usepackage{graphicx}
\usepackage{amsmath}
\usepackage{bm}
\usepackage{xcolor}
\usepackage{xfrac}
\usepackage{arydshln} 
\usepackage{pdflscape} 

\begin{document}





\title{Climate of an Ultra Hot Jupiter}
\subtitle{Spectroscopic phase curve of WASP-18b with HST/WFC3}

\author{Jacob Arcangeli\inst{\ref{inst1}}, Jean-Michel D\'{e}sert\inst{\ref{inst1}}, Vivien Parmentier\inst{\ref{inst2}}, Kevin B. Stevenson\inst{\ref{inst8}}, Jacob L. Bean\inst{\ref{inst3}},  Michael R. Line\inst{\ref{inst7}}, Laura Kreidberg\inst{\ref{inst5}, \ref{inst6}}, Jonathan J Fortney\inst{\ref{inst4}},  Adam P. Showman\inst{\ref{inst9}}}

\institute{Anton Pannekoek Institute for Astronomy, University of Amsterdam, Science Park 904, 1098 XH Amsterdam, The Netherlands \label{inst1} \and
Atmospheric, Ocean, and Planetary Physics, Clarendon Laboratory, Department of Physics, University of Oxford,
Oxford, OX1 3PU, UK \label{inst2} \and 
Space Telescope Science Institute, 3700 San Martin Drive, Baltimore, MD 21218, USA \label{inst8} \and 
Department of Astronomy \& Astrophysics, University of Chicago, 5640 S. Ellis Avenue, Chicago, IL 60637, USA \label{inst3} \and 
School of Earth \& Space Exploration, Arizona State University, Tempe, AZ 85287, USA  \label{inst7} \and 
Center for Astrophysics | Harvard and Smithsonian, Cambridge, MA 02138, USA \label{inst5} \and 
Harvard Society of Fellows, 78 Mt. Auburn St., Cambridge, MA 02138 \label{inst6} \and 
Department of Astronomy and Astrophysics, University of California, Santa Cruz, CA 95064 \label{inst4} \and 
Department of Planetary Sciences and Lunar and Planetary Laboratory, University of Arizona, Tucson, Arizona 85721, USA \label{inst9}}

\abstract{
We present the analysis of a full-orbit, spectroscopic phase curve of the ultra hot Jupiter WASP-18b, obtained with the Wide Field Camera 3 aboard the Hubble Space Telescope. We measure the planet's normalized day-night contrast as >0.96 in luminosity: the disk-integrated dayside emission from the planet is at 964$\pm$25 ppm, corresponding to 2894$\pm30$ K, and we place an upper limit on the nightside emission of <32ppm or 1430K at the 3$\sigma$ level. We also find that the peak of the phase curve exhibits a small, but significant offset in brightness of 4.5$\pm$0.5 degrees eastward. 

We compare the extracted phase curve and phase resolved spectra to 3D Global Circulation Models and find that broadly the data can be well reproduced by some of these models.
We find from this comparison several constraints on the atmospheric properties of the planet. 
Firstly we find that we need efficient drag to explain the very inefficient day-night re-circulation observed. We demonstrate that this drag could be due to Lorentz-force drag by a magnetic field as weak as 10 Gauss. Secondly, we show that a high metallicity is not required to match the large day-night temperature contrast. In fact, the effect of metallicity on the phase curve is different from cooler gas-giant counterparts, due to the high-temperature chemistry in WASP-18b's atmosphere.
Additionally, we compare the current UHJ spectroscopic phase curves, WASP-18b and WASP-103b, and show that these two planets provide a consistent picture with remarkable similarities in their measured and inferred properties. However, key differences in these properties, such as their brightness offsets and radius anomalies, suggest that UHJ could be used to separate between competing theories for the inflation of gas-giant planets.
}

\titlerunning{Phase curve of WASP-18b with HST/WFC3}
\authorrunning{Arcangeli, D\'{e}sert, Parmentier et al.}
\maketitle

\section{Introduction}

Ultra hot Jupiters (UHJs) are gas giants on short orbital periods, typically around early type stars, with dayside temperatures of 2500 K or more. Bright star surveys, such as WASP \citep{Pollacco2006}, KELT \citep{Pepper2007, Pepper2012}, MASCARA \citep{Snellen2013}, are specialised for finding these planets as they are some of the best targets for testing atmospheric theories. This is principally because their high temperatures makes them ideal targets for atmospheric spectroscopy. They also make for convenient chemical laboratories as all of their atmospheric constituents are expected to be in gas phase on their daysides \citep{Parmentier2016}. 

Recent works in the near infrared, supported by observations with the Hubble Space Telescope (HST), have identified the key physics and chemistry that operate at these high temperatures, which can bias retrievals that have been honed on cooler planets. In particular, the dissociation of molecules at low pressures and high temperatures, as well as the opacity of H$^-$ and other molecules strongly influence UHJ spectra (\citealt{Bell2017, Arcangeli2018, Kitzmann2018, Kreidberg2018, Lothringer2018, Mansfield2018, Parmentier2018}). 

UHJs are expected to be tidally locked, ensuring that their daysides are always heated by their host star while their nightsides are permanently dark.
Dayside emission spectra have shown that the majority of the incoming stellar flux must be re-emitted from the daysides of these planets, rather than re-distributed to the nightsides or reflected (e.g. \citealt{Charbonneau2005, Deming2005, Desert2011c, Desert2011d}).
However, these inferred dayside properties are not representative of their global atmospheres, and an understanding of these planets requires consideration of their full 3D atmospheres \citep{Line2016, Feng2016}.
To that end, phase-curve observations allow us to resolve the longitudinal variation in temperature on a planet, and constrain its atmospheric circulation \citep{Knutson2007, Borucki2009, Snellen2009}.
Spectroscopic phase curves, with instruments such as HST, allow us to further break the degeneracies between composition and temperature-structure from the day to the nightside of hot Jupiters \citep{Stevenson2014, Kreidberg2018}.

Phase-curve observations measure key observables, namely the day-to-night contrast of the planet and the brightness offset from the substellar point, that inform the relative balance of wind recirculation and dayside re-radiation \citep{Cowan2007}.
Circulation in hot Jupiters, to first order, can be seen as a balance of two key timescales. These are the radiative and advective timescales, that between them control how efficiently incident dayside flux can be re-distributed to the nightside of the planets \citep{Showman2002}. 

Large brightness offsets have been observed in thermal phase curves of hot Jupiters, pointing toward longitudinally asymmetric temperature distributions. The majority of these planets have temperatures hotter east of the substellar point (see \citealt{Knutson2012, Cowan2012, Parmentier2017}). These eastward offsets are attributed to fast equatorial winds, and are reproduced in first order by Global Circulation Models (GCMs). In these models, it is shown that a super-rotating equatorial jet can form in a hot Jupiter atmosphere and efficiently re-circulate energy from the dayside to the nightside \citep{Showman2002, Showman2011}.  Several planets exhibit westward brightness offsets, the majority of which are dominated by reflected light. Hence these offsets are probing the cloud distribution which is anti-correlated with the temperature map. There are two exceptions to this: HAT-P-7b exhibits a time-variable phase-curve offset \citep{Armstrong2016}, and CoRoT-2b exhibits a strong westward offset in it's Spitzer phase curve \citep{Dang2018}. The time variable offset of HAT-P-7b may be explained by an oscillation of the equatorial wind triggered by MHD effects \citep{Rogers2017}. However, magnetic interactions are unlikely to explain the case of CoRoT-2b \citep{Hindle2019}, whose brightness offset may originate from asynchronous rotation \citep{Rauscher2014}.
The question remains how the observed circulation patterns extend from the classical hot Jupiters to the population of Ultra hot Jupiters, where the additional chemistry and physics that has been identified will influence their circulation.

Due to the high temperatures encountered in UHJ atmospheres, a third timescale is expected to be important, namely the dissipative, or drag, timescale. Strong drag is predicted to occur in the photospheres of UHJ, as their atmospheres should be partially ionized, leading to magnetic braking of waves on the dayside by the planetary magnetic field that acts to impede the formation of an equatorial jet \citep{Perna2010a}. While magnetic braking is not the only source of drag expected to occur in these atmospheres, its strength depends on the ionization fraction and therefore temperature of the atmosphere. Hence the expectation is that highly irradiated objects should have larger day-night contrasts, which is supported by several observations \citep{Komacek2017, Parmentier2017}.

\par
An ideal test case for these theories is the planet WASP-18b \citep{Hellier2009}, which has a high equilibrium temperature of 2413K \citep{Southworth2009} at a period of 0.94 days, placing it firmly in the population of ultra hot Jupiters. In particular, it has a high mass of 10 M$_{Jup}$, occupying the extreme end of the planetary mass regime. This mass places it close to the brown-dwarf regime, who are known to host magnetic fields with a range of field strengths. A planetary magnetic field could have several effects on the planet's observed phase curve and properties: on the day-night contrast through a magnetic drag \citep{Perna2010a}, on the brightness offset through magnetic instabilities \citep{Rogers2014, Dang2018}, or on the radius through Ohmic dissipation \citep{Batygin2010}.

%
In this work, we present an analysis of the spectroscopic phase curve observed with HST Wide Field Camera 3 (WFC3), the third spectroscopic phase curve with HST to be published after WASP-43b and WASP-103b \citep{Stevenson2014, Kreidberg2018}. 
In Section 2 we explain the observations and data reduction methods used. In Section \ref{Sec:Results} we present the results of the white-light phase curve fitting and phase-resolved emission spectra. In Section \ref{Sec:GCMs} we describe the Global Circulation Models used to interpret these results, and the broad properties of WASP-18b that are inferred from this comparison. In Section \ref{Sec:Drag} we further discuss the importance of drag and it's effect on Ohmic dissipation, as well as comparing models with different compositions.  We also place the results of this work in context with with previous analyses of spectroscopic phase-curves, in particular comparing to the UHJ WASP-103b. A final summary of our conclusions is presented in Section \ref{Sec:Conclusions}.

\section{Observations and Data Reduction}

\subsection{Observations}
\par
We observed one phase curve of the hot Jupiter WASP-18b with the Hubble Space Telescope (HST), Cycle 21 Program GO-13467 (PI J. Bean). This phase-curve observation used a total of 18.5 HST orbits over two consecutive visits, covering 2 secondary eclipses and one primary transit of the system.  The data were obtained with the Wide Field Camera 3 (WFC3) aboard HST with the G141 grism, covering 1.1 to 1.7$\mu$m. Details of the observations can be found in \citet{Arcangeli2018} along with a full analysis of the dayside spectrum. In this work we focus on the spectroscopic phase curve. These visits were taken using the 512x512 subarray (SPARS10, NSAMP=15, 112s exposures) in the bi-directional spatial scanning mode, with a half orbit break just before the last secondary eclipse due to a necessary gyro-bias update.
We used our custom data reduction pipeline on the intermediate $ima$ outputs, outlined in \citet{Arcangeli2018}.

\subsection{Systematics correction}

\begin{figure*}[ht]
\includegraphics[scale=0.8]{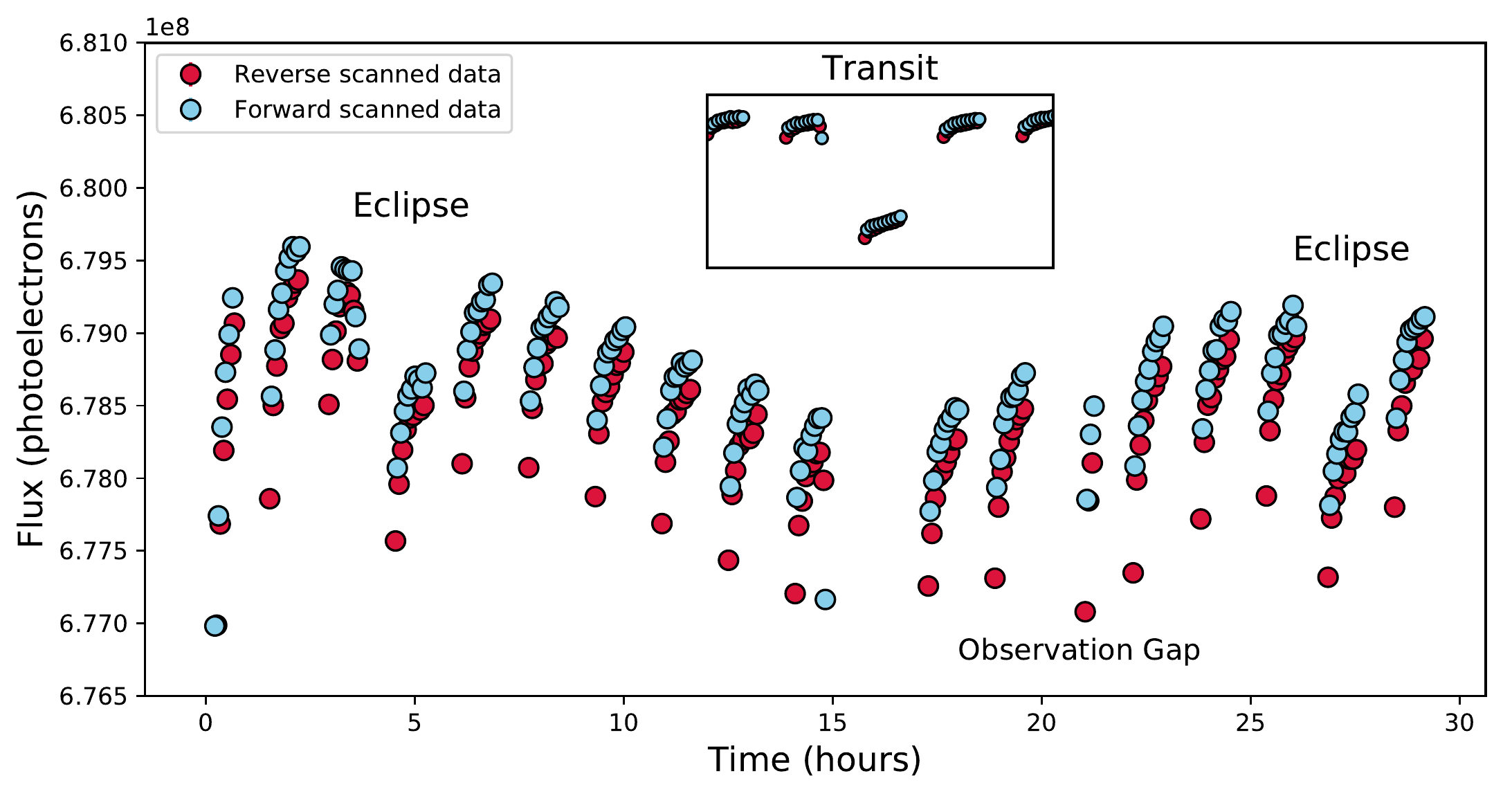}
\caption{Raw observed phase curve of WASP-18b, before removal of systematics and stellar variation. The transit is shown in the inset. The observations cover two eclipses to establish a baseline, with one transit in the middle of the observations. There is an additional half-orbit gap in the observations at 20 hours due to a gyro bias update. Clear orbit-long systematics can be seen, as well as a visit-long slope, on top of the phase-curve variation. The two colours signify the exposures taken from each of the two spatial scan directions, where there is an offset due to the fixed read-out pattern of the detector. }
\label{fig:raw}
\end{figure*}

\begin{figure*}
\includegraphics[scale=0.7]{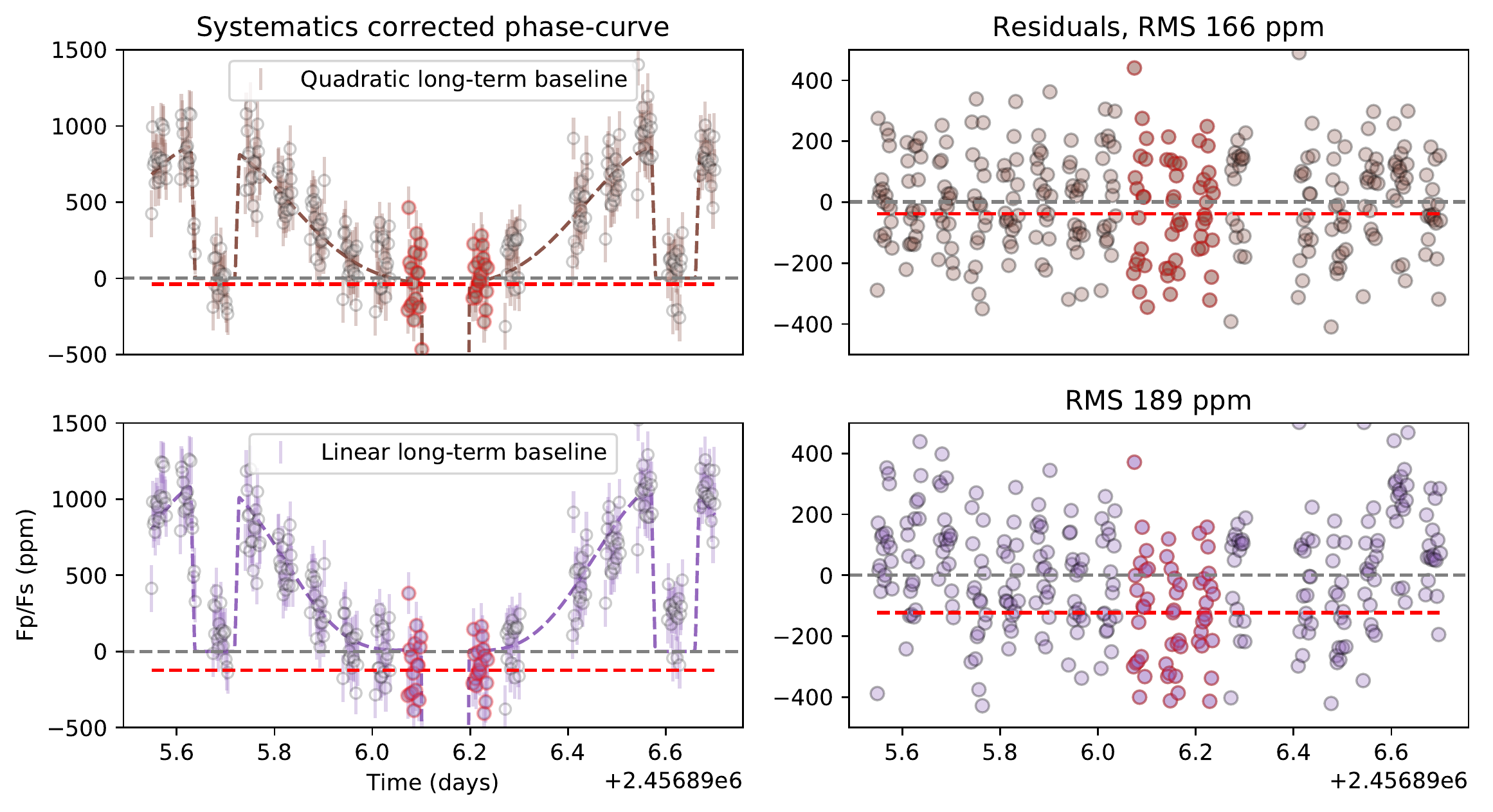}
\caption{Effect of systematic, long-term slope model on extracted phase curve. Left panels: systematics corrected phase curve and best-fit model for the wavelength bin 1.27-1.30 $\mu$m. Right panels: the same residuals plotted in time. The red line is the mean of the nightside residuals. In both panels, the exposures used to calculate the nightside residuals are outlined in red. Using a linear visit-long systematics model degrades the precision of the light curve, introduces visible systematics in the residuals, and results in a significant negative nightside flux. }
\label{fig:vv2}
\end{figure*}

The reduced light curves are dominated by instrument systematics which must be removed in order to extract the planet signal and system parameters (see Figure \ref{fig:raw}). We parametrise the orbit long systematics with a single exponential in time and the visit long systematics with a quadratic function in time, as these have been shown to match well the instrument systematics intrinsic to HST observations \citep{Stevenson2014, Kreidberg2014a}. Consistent with previous analyses, we remove the first orbit of the visit due to the extreme ramp-amplitude, as well as the half-orbit before the second eclipse due to poor sampling of the ramp.
We parametrise the phase curve with a simple two sinusoid model, which is analogous to a Spherical Harmonics model of degree 2. \citet{Kreidberg2018} show that, while the Spherical Harmonics model performs best for the data they analyzed, different phase-curve models can lead to very different temperature maps for the same data. This is due to the intrinsic degeneracy between the measured signals, that can lead to different temperature maps for the same planet \citep{Cowan2012}. 
For the case of WASP-18b, the large mass of the planet should cause significant tidal deformation of the stellar host. The magnitude of this effect is poorly constrained theoretically due to the uncertainty on the stellar density distribution (see Section \ref{Sec:starvar}).
This introduces another degeneracy in the extraction of the planet's signal, hence we choose not to explore different phase-curve models, and only constrain the day-to-night contrast and brightness offset of the planet's phase curve, as they remain consistent between different phase-curve models \citep{Kreidberg2018}. Finally, we do not include the reflected light component in our models, as the albedo of the planet is found to be A$_g<0.057$ \citep{Shporer2018}, which should contribute to <5\% of the total planetary flux at these wavelengths.

We first test our ability to detect the nightside flux in this data by fixing the in-eclipse fluxes to the baseline stellar flux level, and allowing the nightside level to be completely free. We found that we were unable to detect the nightside flux of the planet at a significant level, and in many cases the measured nightside of the planet was below the flux of the star, requiring a negative contribution from the planet which is unphysical (e.g. \citealt{Keating2017}). This is, in part, because the signal of the nightside is expected to be very small compared to the amplitude of the systematics, Fp/Fs<30 ppm for a nightside temperature of 1400K not unexpected for such a system \citep{PerezBecker2013}. Hence, we opted to enforce that the phase-curve model should never fall below the in-eclipse flux, after correcting for systematics and ellipsoidal variations. A stronger constraint would be to enforce that the brightness map of the planet should be non-negative \citep{Keating2018}. Since odd map harmonics discussed in \citet{Keating2018} are not constrained by our data, as they cannot be seen in the phase light curve, and their inclusion or exclusion can change the brightness map of the planet, we opt to only enforce that the phase light curve remain non-negative. We choose not to discuss the brightness map of the planet, as the conversion from phase light curve to brightness map is not unique \citep{Cowan2008}. We verify that, for our best fit parameters, the brightness map can also be made non-negative with the inclusion of odd map harmonics, and hence our results do not require an unphysical brightness map.

In order to reduce the number of degeneracies in our models, we explored the possibility of a linear visit long systematics model rather than a quadratic model. We found that the fit quality of the linear slope model was worse at all wavelengths  when compared to the quadratic model, both with and without a non-negative phase-curve prior. The linear slope also strongly favoured a negative nightside flux when the nightside was set free (shown in Figure \ref{fig:vv2}). Here the quadratic slope model resulted in a nightside flux of -107$\pm$46 ppm (consistent with zero at the 3$\sigma$ level) whereas the linear slope model found a nightside flux of -321$\pm$24 ppm. Finally, the residuals between the best-fit linear-slope model and the data showed clear systematic trends in time, indicating that the model was not fully capturing the data.

\begin{align}
M(t) = & \Big\{ \big[1+ \bm{E_C}\sin(4\pi(\phi(t)-E_\phi)) \nonumber \\
& \qquad + D_C\sin(2\pi(\phi(t)-D_\phi))\big]*T_0(t) \nonumber \\
& + \big[\bm{c_{fp}} + \bm{c_1}\cos(2\pi(\phi(t)-\bm{c_2})) \nonumber \\
& \qquad + \bm{c_3}\cos(4\pi(\phi(t)-\bm{c_4}))\big]*\bm{E_0}(t)\Big\} \nonumber \\ 
& *\bm{C_{scan}}*\big(1+\bm{V_1}t+\bm{V_2}t^2\big)*\big(1-\bm{R_{orb}}e^{-t_{orb}/\bm{\tau}}\big) 
\label{Eq:model}
\end{align}

Equation \ref{Eq:model} shows the full model fitted to the data, where fitted parameters are shown in bold ($\phi$ is the orbital phase of the planet at a given time, $t$ is the time since the beginning of the visit, and $t_{orb}$ is the time since the beginning of an orbit). $T_0(t)$ and $\bm{E_0}(t)$ are the transit and eclipse models respectively, calculated using the batman package \citep{batman}, where we fit for the mid-eclipse time ($\bm{dt1}$) and eclipse depth (through the phase curve parameters). The stellar variations are parametrised by the magnitude of the ellipsoidal variations ({$\bm{E_C}$}), fixed to reach its minimum at transit and eclipse ($E_\phi=0$), while the Doppler boosting signal is fixed to our calculated value of 22 ppm (see Section \ref{Sec:starvar}). The planet signal is modelled by a two-component sinusoid, consisting of 5 free parameters ($\bm{c_{fp}}$, $\bm{c_{1-4}}$). Finally, the instrument systematics are parametrised by the model-ramp approach \citep{Kreidberg2014a, Stevenson2014}. This consists of a quadratic visit-long slope ($\bm{C_{scan}, V_1, V_2}$) and an exponential decay in time for each orbit ($\bm{R_{orb}, \tau}$). $\bm{C_{scan}}$ is the only parameter that is different for each scan direction, and both scan directions are fitted simultaneously. The exponential ramps in each orbit of HST are seen to be stable in time, but are significantly larger in the first orbits of a visit. We therefore allow the first two orbits of the visit, as well as the first orbit after the half-orbit gap, to be fitted with their own ramp amplitudes whilst fixing the other ramp amplitudes to all be equal. We therefore fit for 4 $\bm{R_{orb}}$ for each light-curve ($\bm{R_{1-4}}$).

For each of the wavelength dependent light-curves, the ramp timescale, the ellipsoidal variation, and the eclipse time are fixed to the white-light curve values ($\bm{\tau}, \bm{E_C}, \bm{dt1}$). The remaining parameters are fitted for each channel. We experiment with allowing the ellipsoidal variations to be fitted at each wavelength with a gaussian prior determined by the white-light curve fit, and our results remain consistent. This leads to a total of 16 free parameters for the white-light curve fits and 13 free parameters per spectroscopic channel. 
We fix the remaining system parameters, such as the period and planet/star radius ratio, to values from \citet{Southworth2009}.

The rms of the residuals of the spectroscopic phase-curve fits is between 10-30\% above photon-noise for all bins. The resulting white-light curve fit is shown in Figure \ref{fig:pc} in black and the systematics-corrected data are shown in blue. We find that the residual rms of the best fit to the white-light curve is significantly above photon noise (105 ppm, or 65 ppm above), similar to \citet{Kreidberg2018}. We include a table of our best-fitting parameters in the appendix as well as a corner plot of the posteriors from the white-light curve fit.

\subsection{Estimation of Errors}
In order to estimate the errors on our fitted parameters and identify the degeneracies in the model we use a Markov-Chain Monte Carlo approach using the open-source {\fontfamily{cmr}\selectfont emcee} code \citep{emcee}. Each of the parameters are given flat priors within an acceptable range.
We test the inter and intra chain convergence by employing the Gelmaan-Rubin diagnostic for each of our runs. We run chains of 2,000 steps with 50 walkers to achieve convergence over all parameters.

\section{Results}
\label{Sec:Results}
\subsection{Observed phase-curve properties}

The extracted white-light phase curve shows a large day-to-night contrast, with a peak at 964$\pm$25 ppm just before secondary eclipse and a minimum <32 ppm at 3$\sigma$ level of confidence just before the primary transit (see Figure \ref{fig:pc}). We find that day-night contrast is >0.96 in luminosity, defined here as the difference between the dayside and nightside phase-curve amplitude divided by the dayside amplitude. The peak of the phase curve comes before the secondary eclipse, with a 4.5$\pm$0.5 degree offset of the brightest point eastward in phase. We fit a blackbody for the planet to the dayside spectrum obtained with HST/WFC3, and find a dayside temperature of 2894$\pm$30 K. We place an upper limit on the nightside temperature of 1430~K at the 3$\sigma$ level. For the spectrum of the star, we use the ATLAS9 model atmospheres grid \citep{Castelli2004}, propagating uncertainty on the stellar effective temperature and planet-star radius ratio, taken from \citet{Hellier2009}. These uncertainties on the system parameters correspond to a systematic uncertainty of 25 K on the planet temperature.

These results are consistent with previous phase curves of WASP-18b by \citet{Maxted2013} using Spitzer photometry at 3.6 and 4.5~$\mu$m. They find no evidence of a brightness offset, at a 1$\sigma$ precision of 5 \& 9 degrees in their respective channels, and are unable to detect the nightside of the planet.

We extract the emission spectrum of the planet at different orbital phases (show in Figure \ref{fig:spectra}). We find that the phase-resolved spectra do not exhibit identifiable molecular features of water expected at these wavelengths. The spectra closely resemble blackbody emission at all phases, with decreasing temperature away from the secondary eclipse as expected for a tidally locked planet. This is likely explained by the dayside flux as measured by \citet{Arcangeli2018} dominating the emission spectrum at all phases due to the large day-night luminosity contrast (see also \citealt{Parmentier2018}, Fig. 10).

\begin{table}[ht]
\begin{center}
\begin{tabular}{ | c || c | c |}
\hline
Wavelengths & Eastward Offset & Error \\
$\mu$m & degrees & degrees \\
\hline
\hline
1.14-1.17 & 5.1 & 2.5 \\
1.17-1.21 & 2.8 & 2.6 \\
1.21-1.24 & 3.8 & 2.1 \\
1.24-1.27 & 6.3 & 2.1 \\
1.27-1.30 & 5.4 & 2.0 \\
1.30-1.34 & 6.8 & 2.4 \\
1.34-1.37 & 4.5 & 1.9 \\
1.37-1.40 & 2.5 & 1.7 \\
1.40-1.44 & 6.8 & 2.0 \\
1.44-1.47 & 3.3 & 1.6 \\
1.47-1.50 & 5.8 & 1.7 \\
1.50-1.53 & 5.1 & 1.9 \\
1.53-1.57 & 3.4 & 1.7 \\
1.57-1.60 & 1.6 & 1.3 \\
\hline
1.14-1.60 & 4.5 & 0.5 \\
\hline
\end{tabular}
\end{center}
\caption{Measured brightness offsets for each spectroscopic phase curve. The offsets at each wavelength are consistent with the mean of 4.5$\pm$0.5 degrees to within one or two sigma.}
\label{Tab:Offs}
\end{table}

We additionally measure the offset of the brightest point in the phase curve in each of our 14 wavelength bins (shown in Table \ref{Tab:Offs}) and find that the offset remains constant with wavelength within our uncertainties. We therefore combined the measured offset from each wavelength bin to calculate a white-light curve offset of 4.5$\pm$0.5 degrees eastward in longitude. These brightness offsets are not directly equivalent to hot-spot offsets in the thermal map of the planet, as they measure the offset in integrated hemispheric brightness \citep{Cowan2008, Schwartz2017}. In future discussions we compare only to the brightness offset seen in our data, as the inversion from the light curve to a longitudinal brightness map is not unique.

\subsection{Ellipsoidal variations}
\label{Sec:starvar}
We explored the effects of tidal deformation of the star by the planet and of the planet by the star. Since the mass of the planet is large at 10 M$_{Jup}$, and the planet is only separated by 3.6 Stellar radii from the star, both can have a significant impact on the observed phase curve. 

\begin{figure*}
\includegraphics[scale=0.8]{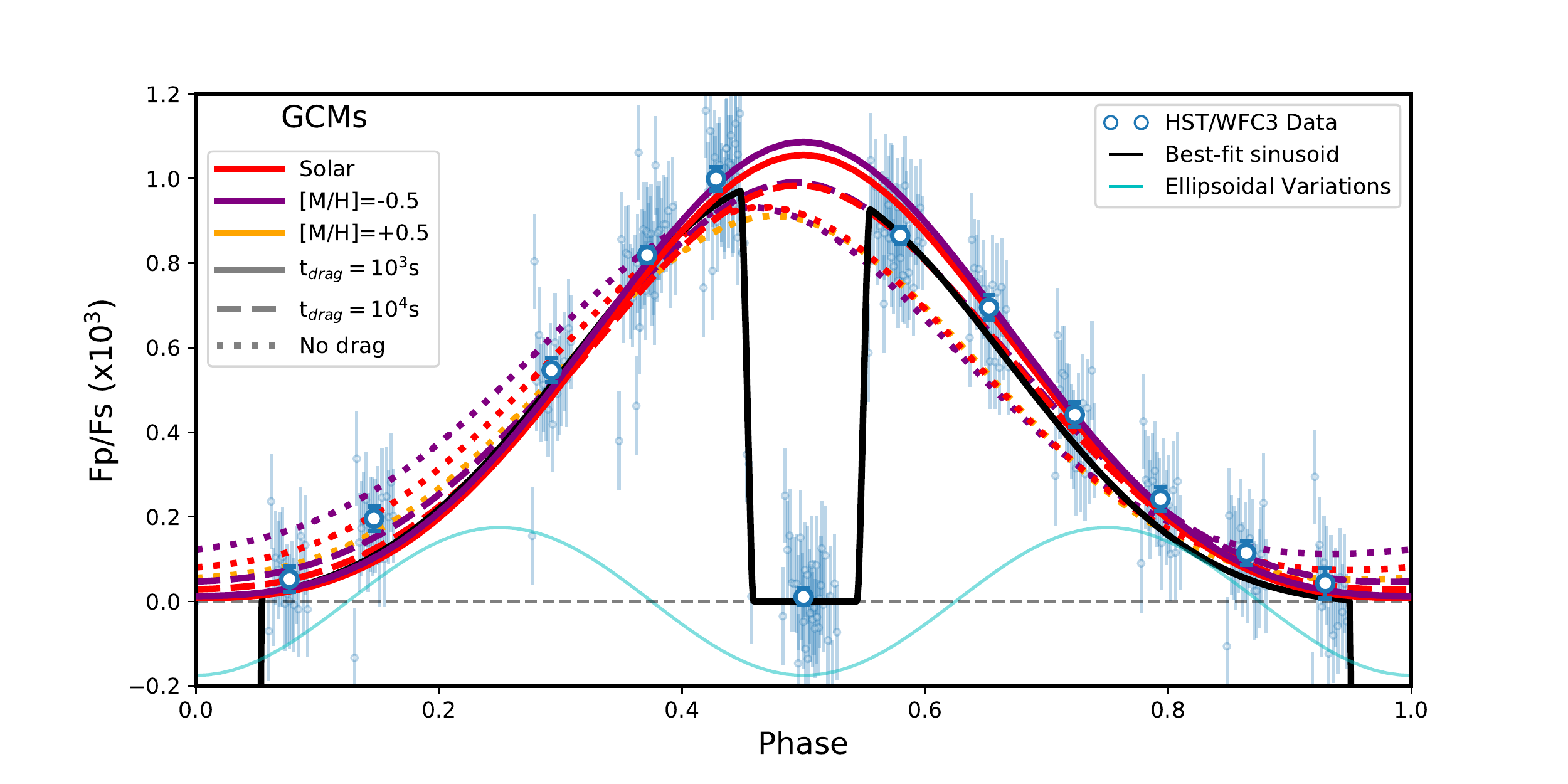}
\caption{Systematics-corrected HST/WFC3 white-light phase curve of WASP-18b (blue points) compared to suite of Global Circulation Models (coloured curves). In black is the best-fit model used to parametrize the planet signal, with two sinusoidal components for the phase-curve variation. In cyan are the stellar ellipsoidal variations, that have been subtracted from the phase curve (see Section~\ref{Sec:starvar}). GCMs shown are parametrized by 3 drag timescales indicated by line-styles and by 3 metallicities indicated by colours (-0.5, 0.0, and +0.5 relative to solar).}
\label{fig:pc}
\end{figure*}

We use the equations supplied in \citet{Leconte2011b} and estimate that the stellar ellipsoidal variations are of order 200 ppm, or 400 ppm peak-to-peak. The size of these variations is uncertain due to the uncertainty on the stellar density distribution.
Since the ellipsoidal variations operate on the same timescale as our phase-curve model, they are a significant source of degeneracy. We include the magnitude of the ellipsoidal variations in our fits and remove them from our final light curves. The fitted magnitude of the stellar ellipsoidal variation is 201$\pm$26 ppm in the white-light curve, with a phase offset fixed such that the minima are at secondary eclipse and transit (shown as the cyan curve in Figure \ref{fig:pc}). We fixed the magnitude of the ellipsoidal variations to the white-light curve values in each of the spectroscopic phase curves.

Interestingly, our measurement of the stellar ellipsoidal variation is consistent with the independent measurement of the ellipsoidal variations by \citet{Shporer2018} to within the 1$\sigma$ confidence level. \citet{Shporer2018} measured the amplitude of the ellipsoidal variations as 194$\pm7$ ppm using observations from the \textit{TESS} spacecraft in the optical. 
Additionally the dayside emission of WASP-18b measured with \textit{TESS} is consistent with our dayside emission spectrum, as discussed in \citet{Shporer2018}. However, while the planetary emission at other phases is consistent within the errorbars of \citet{Shporer2018} and this work, the differences in how the planetary signal is extracted make a comparison difficult. For instance the nightside flux in \textit{TESS} is measured at -24 ppm, whereas our approach enforces that the nightside signal is non-negative.
We do not fix our ellipsoidal variations to the more precise measurement from \citet{Shporer2018} in order to allow us to compare the two results as independent measurements. We do however test our analysis using their value of EC=194$\pm$7ppm, and find our conclusions unchanged. The most significant effect of including their more precise measurement is an increased precision on the phase curve at quadrature, where the ellipsoidal variations peak.

We estimate the inverse effect, the tidal deformation of the planet by the star, using the tables supplied in \citet{Leconte2011a} for a 10M$_J$ planet under high irradiation. We find that the expected size of the planet's variations are only an effective 0.25\% in radius, equivalent to 5ppm in the final light curve, which should be negligible at our precision and is well within the errors on the measured radius of the planet of $\pm6\%$ \citep{Southworth2009}.
We also estimate the effect of doppler boosting due to the radial velocity of the star on these light-curves to be about 22 ppm \citep{Mazeh2010}, and include it in our models fixed to this value. This is also consistent with the measurement of 24$\pm$6 ppm by \citet{Shporer2018}.

An additional effect of the ellipsoidal variations is that they may offset any measurements of the transit or eclipse depths when unaccounted for (e.g. \citealt{Cowan2012}). This effect is mitigated in the case of WASP-18b, as the planet's phase-curve variation within the HST/WFC3 bandpass is by coincidence nearly equal in magnitude to the ellipsoidal variations over the duration of the eclipse. We estimate the difference in the retrieved eclipse depths, when the phase-curve and ellipsoidal variations are fixed to our best-fit values versus when they are unaccounted for, and modelled as instrument systematics. When these effects are not accounted for, we find a relative change in eclipse depth of only 3 ppm over the whole spectrum, well below the precision of the data, and a systematic over-estimate of the eclipse depths (offset) of about 20 ppm or $\sim$2\% of the depth. These changes are within the 1-sigma errors of Arcangeli+2018. However, we stress that for other planets where these effects do not necessarily cancel, it has been shown that their treatment can significantly affect inferred eclipse depths \citep{Cowan2012, Kreidberg2018}.

\section{Comparing the phase curve of WASP-18b to Global Circulation Models}
\label{Sec:GCMs}

\begin{figure*}
\includegraphics[scale=0.58]{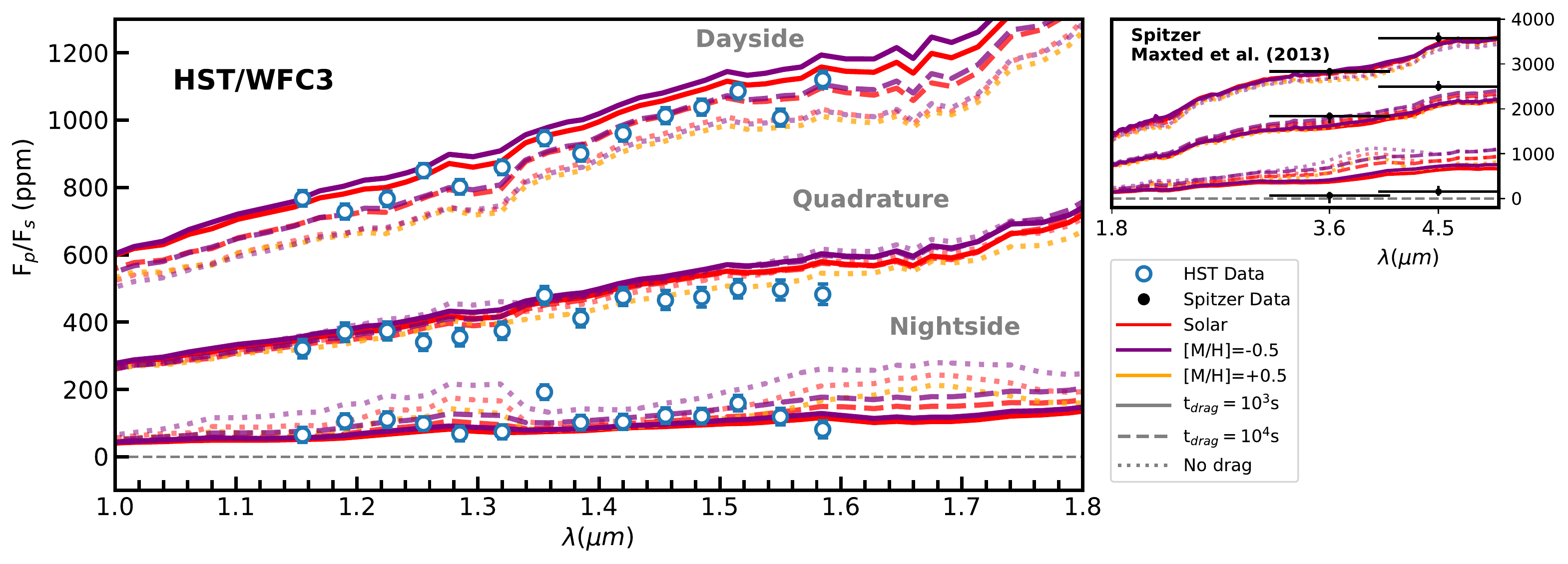}
\caption{ HST/WFC3 phase-resolved spectra at three different phases (dayside, nightside, and quadratures) compared to range of GCMs. Spitzer points are taken from the best-fit sinusoidal phase-curve model by \citet{Maxted2013}, where the nightside was fixed to zero flux, with errorbars taken from their eclipse depth measurements.
The phase bins shown are three broad bins: "Dayside" refers to orbits between phases 0.4-0.45 and 0.55-0.6, "Quadrature" to the average of phases 0.25-0.3 and 0.7-0.75, and "Nightside" to phases 0.05-0.25 and 0.75-0.95. The same binning is used for the GCM spectra shown.}
\label{fig:spectra}
\end{figure*}

\subsection{Global Circulation Models}

To interpret our data and determine the physical origin of the observed signals, we compare the extracted phase curve and spectra to 3D Global Circulation Models (GCMs). We produced a sample of circulation models, exploring the effects of changing drag and metallicity as these have been shown to determine the broad behaviour of hot Jupiter phase curves \citep{Showman2011,Kataria2015}. Here drag refers to any dissipative mechanism that can act to slow down wave propagation and reduce windspeeds.

The atmospheric circulation and thermal structure were simulated using the SPARC/MITgcm model \citep{Showman2009}. The model solves the primitive equations in spherical geometry using the MITgcm \citep{Adcroft2004} and the radiative transfer equations using a state-of-the-art one dimensional radiative transfer model \citep{Marley1999}. We use the correlated-k framework to generate opacities, based on the line-by-line opacities described in \citet{Visscher2006, Freedman2014}. Our initial model assumes a solar composition with elemental abundances of \citet{Lodders2002} and the chemical equilibrium gas phase composition from \citet{Visscher2006}. These calculations take into account the presence of H$^-$ opacities and the effect of molecular dissociation on the abundances, shown to be important for this class of planet \citep{Arcangeli2018, Kreidberg2018, Mansfield2018, Parmentier2018, Bell2018}. Additional heat transport by H$_2$ recombination is not included in our models \citep{Bell2018}. We used a timestep of 25s, ran the simulations for 300 Earth days, averaging all quantities over the last 100 days. The above modelling process is the same as that described in \citet{Parmentier2018}, using the WASP-18 system parameters from \citet{Southworth2009}.

We include additional sources of drag through a Rayleigh-drag parametrisation with a single constant timescale per model that determines the efficiency with which the flow is damped. We vary this timescale between models from $t_{drag}=10^{3-6}s$ (efficient drag), as well as a no drag model with $t_{drag}=\infty$. While all the models are radiatively dominated on the dayside, our range of drag strengths cover the transition from a drag-free, wind circulation case to a drag-dominated circulation.
This can be seen from the short radiative timescale of the dayside photospheres of our models, $t_{rad}\sim10^{2-3}$s, estimated using Eq. 10 from \citet{Showman2002} for a simple H$_2$ slab atmosphere. This is significantly shorter than other relevant timescales, such as the advective timescale at the equator, calculated as the ratio of the equatorial windspeed over the planet radius. The advective timescale is on the order of $10^4$s in our no drag model rising to $10^6$s in our efficient drag model ($t_{drag}=10^3$s), as the model atmospheres transition to drag-dominated circulations.

\subsection{Comparison of GCMs to data}

To first order, all of our Global Circulation Models show a large day-night contrast and small or no brightness offset, and broadly reproduce the observed phase curve of WASP-18b.
We find however that our baseline, solar-composition model with no additional drag sources fails to match the size of the day-night contrast in the phase-curve data (Figure~\ref{fig:pc}). This baseline model both under-predicts the dayside flux and over-predicts the nightside flux of the planet.
However, models with additional sources of drag, parametrised by a short drag timescale, are able to match better the dayside and nightside flux of the planet.
All of our efficient drag models match well the day-night contrast of the planet. We therefore find that we require additional drag sources to explain our observed day-night contrast of WASP-18b. We discuss the possible physical origins of this additional drag in Section~\ref{Sec:Magnets}.

When generating our GCMs, we held the system parameters fixed to literature values. We were not able to explore their full uncertainties as generating one model is already computationally expensive. However, uncertainties on the system parameters will lead to an additional uncertainties on the inferred best-fit models. Since the differences between the efficient drag models are so small, we cannot statistically choose between these models (see Table~\ref{Tab:gcms}).
Nevertheless, models with inefficient or no drag fail to reproduce the day-night contrast seen in the data and therefore the overall shape of the phase curve. 
Hence we can conclude that we require some additional source of efficient drag to explain our observed phase curve, and the best-fitting models are those with $t_{drag}=10^3$s or $10^4$s.

We test whether enhanced or depleted metallicity might also explain the large day-night contrast without the need for additional drag sources \citep{Kataria2015}. We find however that the effect of changing metallicity alone, in our models of WASP-18b, is too small to explain the observations (see Figure~\ref{fig:pc}), but the effect of metallicity on the phase curve is different for these hot planets compared too cooler objects (see Section~\ref{Sec:MH}).


We also compare our GCMs to phase-resolved spectra extracted from the spectroscopic phase curve, as this allows us to study the wavelength dependence of our data. As seen in Figure \ref{fig:spectra}, we find that the spectra are best matched by the efficient drag models. The phase-resolved spectra are dominated by thermal emission from the dayside due to the large day-night luminosity contrast.

\renewcommand{\arraystretch}{1.3}
\begin{table}[ht]
\begin{center}
\begin{tabular}{ | c | c | c | c |}
\hline
Model & Model & $\Delta\chi^2$ & $\Delta\chi^2$ \\
 t$_{drag}$ (s) & [M/H] &   & per datum \\
\hline
$10^3$ & -0.5 &  37 & 0.2 \\
$10^4$ & -0.5 &  0 & 0.0\\
no drag & -0.5 & 137 & 0.6 \\
$10^3$ & 0.0 &  23 & 0.1 \\
$10^4$ & 0.0 &  7 & 0.0 \\
no drag & 0.0 &  73 & 0.3 \\
no drag & +0.5 & 70 & 0.3\\
\hline
\end{tabular}
\end{center}
\caption{Table of the fit quality between our circulation models and the measured phase curve of WASP-18b, relative to the best fitting model. Fit quality here is measured by the $\chi^2$ between the data and the circulation model.}
\label{Tab:gcms}
\end{table}


\section{Discussion}
\label{Sec:Drag}
\subsection{Drag in Global Circulation Models} 
\label{Sec:Magnets}
\subsubsection{Sources of drag}

We infer the presence of an efficient drag source in the atmosphere of WASP-18b from our comparison to Global Circulation Models. We explore the origins of this drag, which could originate from a variety of sources, such as turbulence and instabilities \citep{Goodman2009,Li2010, Youdin2010}, or hydrodynamic shocks \citep{DobbsDixon2010, Rauscher2010, Heng2012, Fromang2016}. Alternatively, for hot planets whose atmospheres are partially ionized, magnetic fields may influence the circulation and create a magnetic drag \citep{Perna2010a, Batygin2013, Komacek2017}.
Magnetic drag, sometimes called "ion drag", is caused by the collision between the bulk neutral flow and the ionic component of the flow. Since the ionic component is subject to Lorentz forces but the neutral component is not, these ions can act as a drag and eventually dominate the circulation under the right conditions \citep{Zhu2005}.

Alkali metals in the dayside atmosphere of WASP-18b should be significantly thermally ionized \citep{Arcangeli2018, Helling2019}. Thus, if WASP-18b were to host a magnetic field, its circulation would be influenced by magnetic drag.
We use the formula described in \citealt{Perna2010a} (here Equation~\ref{Eq:drag}) to estimate what might be the efficiency of magnetic drag on this planet, parametrised through the timescale t$_{drag}$.

\begin{align}
\qquad \qquad \qquad \qquad t_{drag} \sim \frac{4\pi\rho\cdot\eta(n_e)}{B^2cos\theta} \quad \text{Perna et al. (2010a)}
\label{Eq:drag}
\end{align}

This timescale t$_{drag}$ is the timescale on which kinetic energy is dissipated by magnetic drag in the atmosphere. We calculated the ionization fraction in local chemical equilibrium for our circulation models with a modified version of the NASA CEA Gibbs minimization code~\citep[e.g.][]{Gordon1994,Parmentier2018}. This ionization fraction ($n_e$) is used to compute $\eta$, the resistivity, defined in \citet{Perna2010a}. B here is the magnetic field strength, $\rho$ is the density of the atmosphere, and $\pi-\theta$ is the angle between the magnetic field and the flow (assumed here that $\theta=0$ for the case of maximal efficiency).

We find that we can match a t$_{drag}$ of order $10^3$s at the substellar point, rising to $10^4$s elsewhere on the dayside, with a magnetic field strength of B=10 Gauss . The nightside drag timescale from our models is much longer, as the nightside is too cold for thermal ionization. This estimate shows that the short drag timescale inferred from our GCMs could reasonably be due to Lorentz forces. However, the inhomogeneity also highlights that our Rayleigh-drag parametrisation does not capture the full effects of magnetic drag, as for instance a magnetic drag timescale should vary throughout the atmosphere.

\subsubsection{Limitations of Rayleigh drag parametrisation}

In our GCMs, drag is parametrized by a simple Rayleigh drag \citep{Komacek2016} which aims to approximate any additional drag sources by one single dissipative timescale throughout the planet, without the need for finer resolutions or full MHD calculations. This parametrization is shown in Equation \ref{Eq:fdrag}, where the drag force per unit mass, $\mathcal{F}_{drag}$, is given as $\textit{\textbf{v}}$, the velocity of the flow, divided by $t_{drag}$, a single constant drag timescale. 

\begin{align}
\qquad \qquad \qquad \qquad \quad \mathcal{F}_{drag} = -\frac{\textit{\textbf{v}}}{t_{drag}} 
\label{Eq:fdrag}
\end{align}

In the context of magnetic drag, this parametrization has three key problems. The first is that the magnetic drag is direction-dependent with respect to the magnetic field \citep{Batygin2013} seen as the $\theta$ term in Equation \ref{Eq:drag}. The second is that the strength of drag is spatially inhomogeneous, i.e. it is weaker in cooler regions where there is less ionization, such as the nightside hemisphere \citep{Rauscher2012}.
Finally, the atmospheric circulation itself can induce a toroidal magnetic field that is larger in amplitude than the dipole field of the planet, that can change both the strength and the direction of the Lorentz force~\citep{Rogers2014}.
All these aforementioned effects are not accounted for in our models which could limit our accuracy in predicting the shape of the phase curve.

\begin{figure}
\includegraphics[scale=0.48]{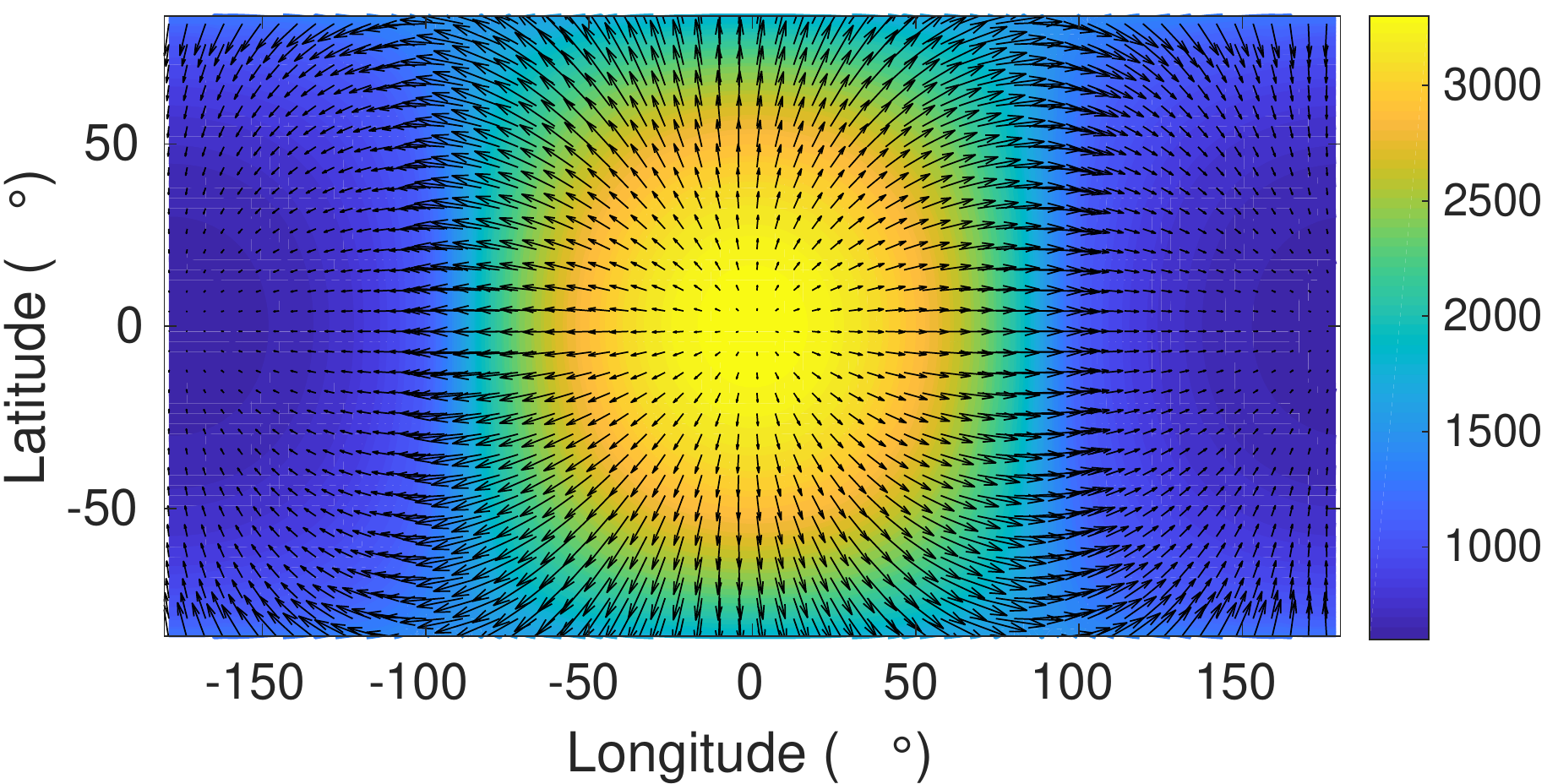}
\includegraphics[scale=0.48]{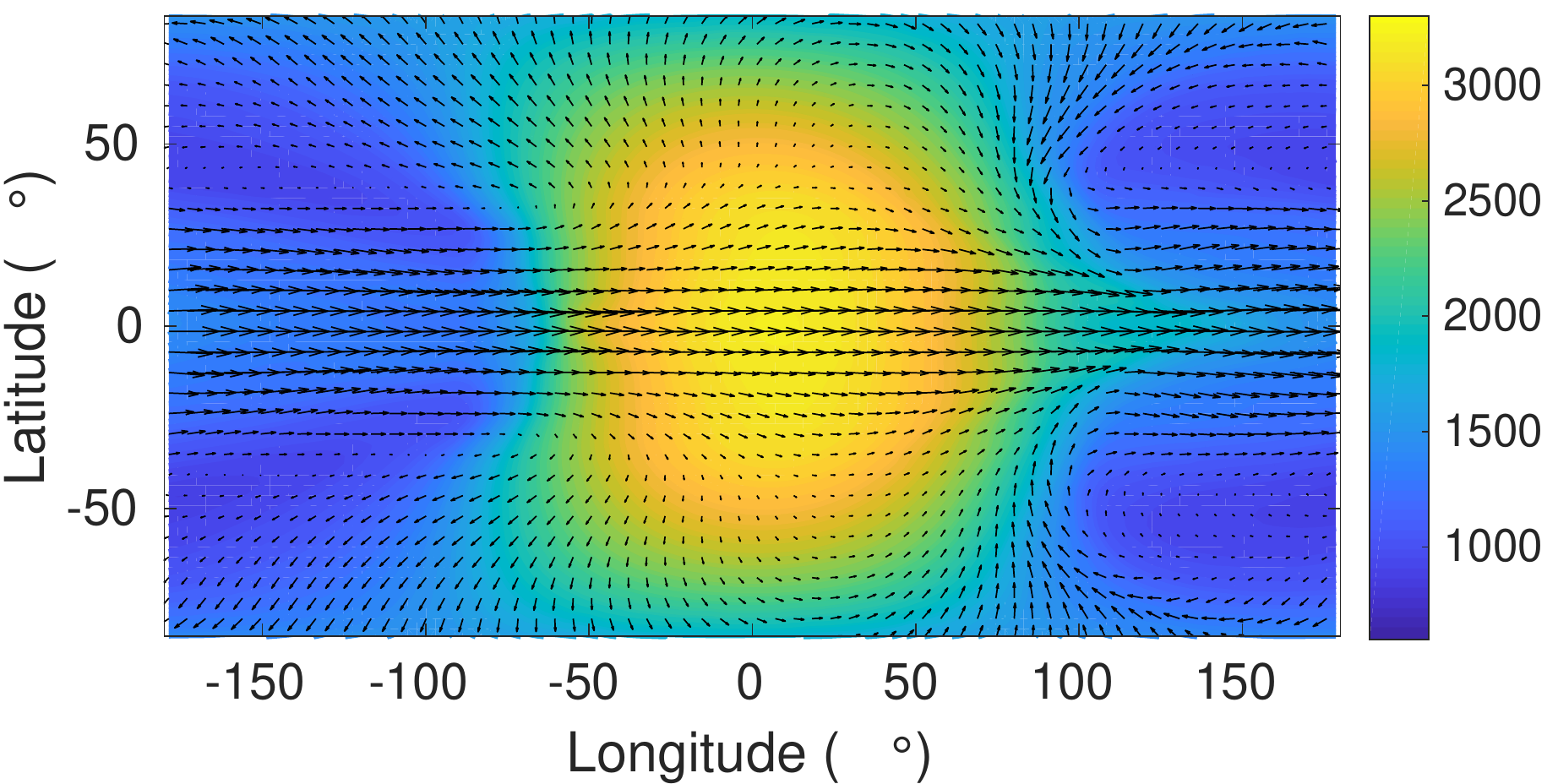}
\caption{Temperature and wind maps of WASP-18b taken from GCMs at a pressure level of 0.2 bar for two cases of drag. \newline \textbf{Top:} Efficient drag model (t$_{drag}=10^3$s), where the circulation is day-to-night and longitudinally symmetric. \newline \textbf{Bottom:} No drag model, the circulation to the nightside is more efficient and driven by an equatorial jet.}
\label{fig:circ}
\end{figure}

\subsection{Effect of a planetary magnetic field}

\subsubsection{Magnetic Circulation}
\citet{Batygin2013} explored further the effect of the directionally-dependent Lorentz-force on atmospheric circulation. Our GCMs with very efficient drag exhibit longitudinally symmetric day-night flow patterns (top panel of Figure~\ref{fig:circ}), as efficient drag shuts down Rossby and Kelvin waves in the dayside atmosphere that are responsible for the typical equatorial jet formation \citep{Showman2011}.
However \citet{Batygin2013} suggest that this should only occur on objects with low magnetic fields, typically with B<0.5 Gauss in strength. They predict that the majority of highly irradiated gas giant atmospheres should be dominated by zonal jets \citep{Showman2011}, such as those seen in our drag free models (bottom panel of Figure~\ref{fig:circ}). This is due to Lorentz-forces that act perpendicular to the magnetic field lines leading to circulation patterns with zonal jets, rather than simply damping the existing flow as in the Rayleignash-drag parametrization \citep{Batygin2013}.
In contrast, without considering magnetic effects, drag due to other schemes such as the Kelvin-Helmholtz instability should lead to a symmetric day-night flow with no brightness offset.
In this context, the presence of a brightness offset in our data might be an indication that magnetic effects are responsible for the observed circulation on WASP-18b, and therefore favour a magnetic drag scenario. 

Magnetic drag should affect the circulation of cooler planets more than the Ultra hot Jupiters. As discussed in \citet{Perna2010a} for HD209458b ($T_{eq}$=1460K), magnetic drag may act to limit the wind speeds, but will not completely shut down re-circulation for magnetic fields with B<30 Gauss or so. This is because there is a direct dependence of the magnetic drag timescale on the resistivity of the atmosphere (see Equation~\ref{Eq:drag}). This resistivity depends strongly on the local temperature through the ionization fraction, which drops sharply even across the dayside of WASP-18b (see \citealt{Helling2019}, Figure 6).

\subsubsection{Ohmic dissipation in WASP-18b}
\label{Sec:Inflation}

Another consequence of a planetary magnetic field is Ohmic dissipation which has been proposed as a mechanism to explain hot Jupiter inflation \citep{Batygin2010, Perna2010b, Menou2012}. Here we consider what the effect of Ohmic dissipation would be on the measured radius of WASP-18b.

In the specific case of WASP-18b, there are two additional factors affecting the radius inflation by Ohmic dissipation due its high temperature and large mass. The first is that heating due to Ohmic dissipation should be less efficient in the presence of efficient drag, as the zonal winds that drive Ohmic dissipation are reduced in speed \citep{Menou2012}. The second is that inflation by Ohmic dissipation should be less efficient for higher mass planets \citep{Huang2012}. This is because the depth at which Ohmic power is deposited depends on the scale-height of the atmosphere, and energy should be deposited higher in massive planet atmospheres than less massive planets. This would reduce the effect of any additional heating by Ohmic dissipation on the observed radius of massive planets.

These effects combined predict that WASP-18b should not be significantly inflated by Ohmic dissipation.
Using predictions from \citet{Thorngren2018}, a 10 M$_{J}$ planet should have a maximum radius of 1.21 R$_{J}$ when inflation effects are not taken into account. The measured radius of WASP-18b is consistent with this value within one sigma (1.204$\pm$0.035 \citealt{Maxted2013}). The effect of inflation on the radius of high mass planets is harder to detect \citep{Miller2009}, however we can conclude that WASP-18b does not deviate from the scenario predicted by Ohmic dissipation. In Section \ref{Sec:Comparison}, we show that we can lift some of this degeneracy between mass and inflation efficiency by comparing to a lower mass UHJ, WASP-103b.


\subsection{Effect of Atmospheric Metallicity on the phase curve of WASP-18b}
\label{Sec:MH}

Previous work has shown that changes in atmospheric metallicity can affect global circulation in hot Jupiter atmospheres \citep{Showman2009, Kataria2015}. 
For instance, increasing the metallicity in a classical hot Jupiter model typically acts to reduce the planet's nightside emission (e.g. WASP-43b, \citealt{Kataria2015}). This is because, as the metallicity increases, the abundances of molecular species increase leading to stellar light being absorbed higher in the atmosphere, at lower pressures.
The radiative timescale is shorter at lower pressures, leading to less efficient heat circulation to the nightside of the planet \citep{Showman2009}. Additionally, as more incident energy is re-irradiated from the dayside, the brightness temperature of the dayside increases with increasing metallicity for a classical hot Jupiter model \citep{Kataria2015}.

WASP-18b resides in a hotter regime, where for instance gas phase TiO becomes an important chemical species.
For WASP-18b, an increase in atmospheric metallicity changes the abundance ratio of gas phase TiO vs H$_2$O above the photosphere \citep{Parmentier2018}. While TiO absorption increases the temperature above the band-averaged WFC3 photosphere, the temperature at the WFC3 photosphere decreases, as correspondingly less stellar light reaches photospheric pressures. 
Hence the planet's dayside is dimmer in the WFC3 bandpass in the higher metallicity case, but should be brighter in emission shorter than $1\mu m$, inside the TiO bandpass.
The reverse effect is seen when the metallicity is decreased: the temperature of the WFC3 photosphere is hotter but the nightside re-circulation is more efficient, leading to an increase in emission at all phases in the WFC3 bandpass.

Neither an enhancement or depletion of $\pm$0.5 in metallicity relative to solar is sufficient to match the observed phase curve of WASP-18b without including efficient drag in our models. However, there is a strong dependence of the nightside temperature with metallicity in our no-drag models (see Figures~\ref{fig:pc} and \ref{fig:spectra}).\par


\subsection{Constraints on the Redistribution}

\begin{figure}[h]
\includegraphics[scale=0.5]{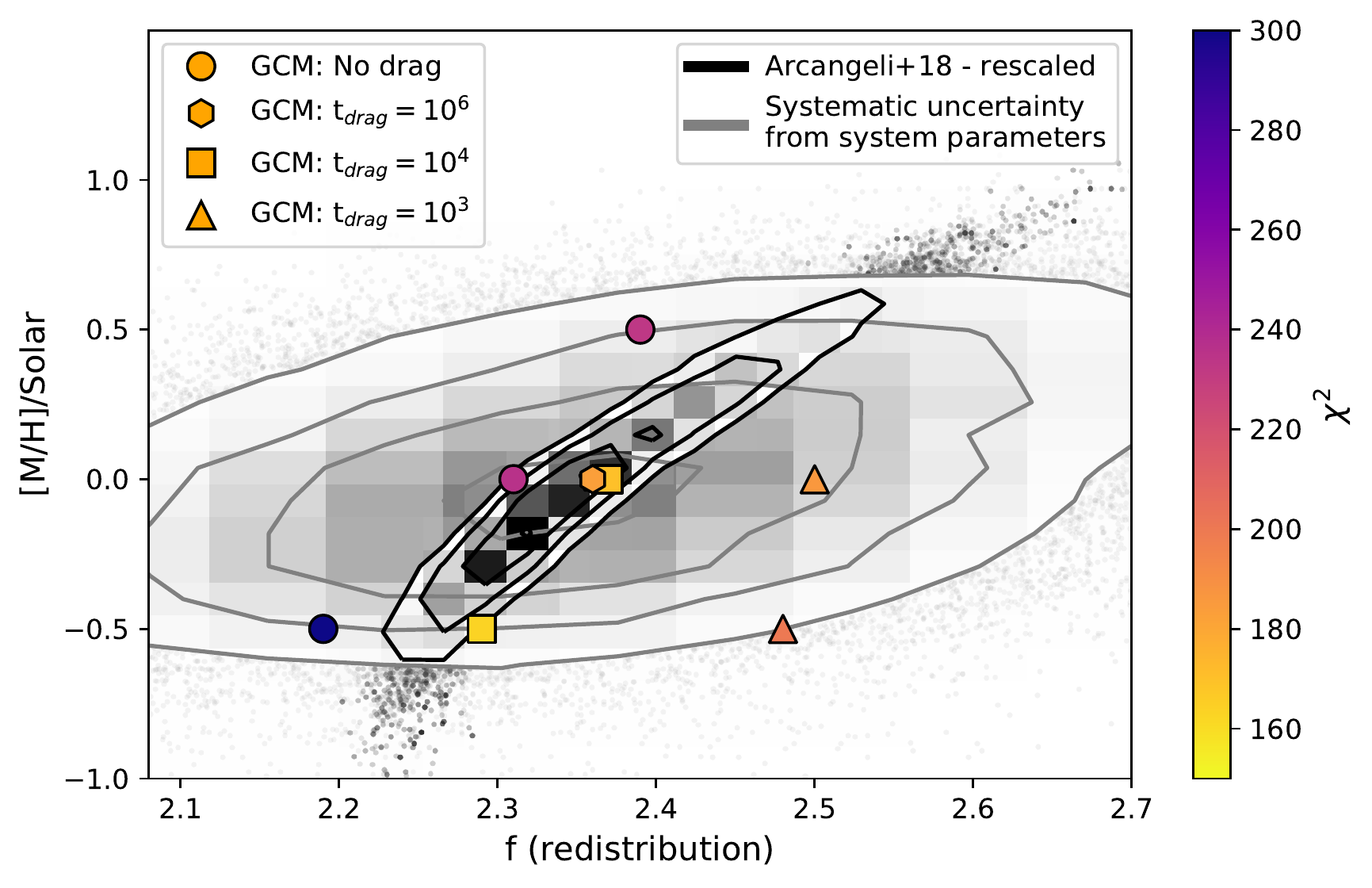}
\caption{Metallicity as a function of redistribution efficiency, comparing 1D modelling to our 3D GCMs. GCM outputs are shown by coloured markers, where marker styles indicate the drag strength in each model: from right to left t$_{drag}$=$10^3$s (triangles), $10^4$s (squares), $10^6$s (hexagon), and no-drag (circles). Markers are also coloured by $\chi^2$ between the corresponding GCM phase curve and our HST/WFC3 phase curve.
Retrieved redistribution efficiencies from \citet{Arcangeli2018} are shown by black contours, while in grey are the same contours including a systematic uncertainty on the equilibrium temperature of 44~K.}
\label{fig:f}
\end{figure}

The redistribution efficiency determines the amount of flux that is re-emitted from the dayside versus the amount of flux that is carried to the nightside. It can be defined through the ratio of the dayside temperature and the equilibrium temperature of the planet: $f=(T_d/T_{eq})^4$. Here $f=2$ refers to the dayside-only redistribution case, while $f=2.67$ refers to the no-redistribution case (where the dayside reaches the maximum temperature).
The redistribution efficiency can be used as a simple measure of the atmospheric circulation regime.
We show the redistribution efficiencies for each of our GCMs in Figure~\ref{fig:f} as the coloured points. All of our GCMs have redistribution factors between $2.2<f<2.5$.

These models for WASP-18b show that the redistribution efficiency is strongly dependent on metallicity in the case of no or weak drag. This can be seen as the steep dependence of redistribution efficiency with metallicity for the left-most points in Figure~\ref{fig:f} (the weak drag regime). The origin of this effect is described in Section~\ref{Sec:MH}.
For models with efficient drag, the dependence of redistribution on metallicity is greatly reduced. In these models the circulation becomes drag dominated and inefficient at all pressures, hence the effect of metallicity is less pronounced.

Typically the redistribution efficiency cannot be estimated solely from the dayside emission, as there is a known degeneracy between the albedo and the redistribution efficiency \citep{Cowan2011}. For UHJs, while the albedos are expected to be very small, we illustrate that the uncertainty on the equilibrium temperature limits any conclusions that could be drawn from the dayside alone. The redistribution efficiencies calculated in \citet{Arcangeli2018} adopted fixed stellar parameters during the retrieval process, fixing $T_{eq}=2477$ K. This is offset from our GCMs, calculated using slightly different stellar parameters, leading to $T_{eq}=2385$ K. We correct for this offset and include a systematic uncertainty of $\Delta f=\pm0.18$ from $\Delta T_{eq}=\pm44K$ \citep{Southworth2009}.  In Figure~\ref{fig:f}, we plot these modified metallicity/redistribution contours from \citet{Arcangeli2018}. Here we see that the uncertainty on $T_{eq}$ dominates over the retrieved errors, and needs to be included when retrieving the redistribution from the dayside alone.

\subsection{Comparison of two Ultra hot Jupiters}
\label{Sec:Comparison}

Spectroscopic phase curves with HST/WFC3 have been published for 3 exoplanets so far, WASP-43b \citep{Stevenson2014}, WASP-103b \citep{Kreidberg2018}, and WASP-18 (this work). Of these planets, WASP-18b and WASP-103b both belong to the class of Ultra hot Jupiters, while WASP-43b is in a cooler regime (T$_{eq}=1370$ K).
The inferred properties, spectra, and phase curves of these two UHJs have many similarities (see Table \ref{Tab:W103}). However, key differences between these planets remain, namely their masses, their radius anomalies, and their measured brightness offsets.

\begin{table}[ht]
\begin{center}
\begin{tabular}{ | c || c | c |}
\multicolumn{3}{c}{\textit{System Parameters}}  \\
\hline
System & WASP-18 & WASP-103 \\
\hline
\hline
Planet Mass $M_{J}$  & 10.43$\pm0.54$  & 1.49$\pm$0.09  \\
Planet Radius $R_{J}$ & 1.17$\pm0.07$ & 1.55$\pm$0.05 \\
Planet T$_{eq}$ & 2413$\pm44$ K & 2508$\pm73$ K  \\
Orbital Period & 22.6 h & 22.2 h \\
Stellar T$_{eff}$ & 6400$\pm100$ K & 6110$\pm$160 K \\
\hline
\multicolumn{3}{c}{\textit{Inferred Properties}}  \\
\hline
Day-night contrast & $\bm{>0.96}$ & 0.93  \\
Brightness offset & $\bm{-4.5^\circ\pm0.5}$ & -0.3$^\circ\pm$0.1  \\
Planet Metallicity & $-0.01\pm0.35$ & $1.36\pm0.36$ \\
Dayside Photosphere & 0.33 bar & 0.01 bar \\
Radius Anomaly & 0\% & 41\% \\
\hline
\end{tabular}
\end{center}
\caption{Comparison of measured and inferred properties of the WASP-18 and WASP-103 Ultra hot Jupiter systems. System parameters are taken from \citet{Hellier2009, Southworth2009}; and \citet{Southworth2015}. Values in bold are from this work. Other inferred properties are taken from \citet{Arcangeli2018, Kreidberg2018}. Photospheric pressures here are the median contributing pressures of the dayside spectra over the HST/WFC3 G141 bandpass. The radius anomaly shown is the difference in radius between the measured value and the non-inflated model as a percentage of the measured radius taken from \citet{Thorngren2018}. }
\label{Tab:W103}
\end{table}

The differences between these two planets can shed further light on their atmospheric properties. A first major difference is in their observed circulations: WASP-18b has a small but significant phase-curve offset whereas the phase curve of WASP-103b appears longitudinally symmetric.
In the simple picture of circulation as a balance of the radiative and advective timescales \citep{Showman2002}, we would expect the day-night contrast of WASP-18b to be lower than WASP-103b, as the brightness offset should correspond to moderately efficient wind-driven circulation.
However the observed day-night contrast of WASP-18b is larger than WASP-103b, in conflict with this simple picture. As suggested in Section \ref{Sec:Magnets}, the observed offset of WASP-18b may be the result of a magnetic field, and an offset may not be present on WASP-103b if the planet were to host a weaker magnetic field, as might be expected from its lower mass \citep{Yadav2017}.

The second major difference is that WASP-103b is very inflated, while WASP-18b is consistent with a non-inflated model (see Section \ref{Sec:Inflation}). The radius of WASP-103b should be about 1.10 R$_{Jup}$ when no inflation mechanism is present \citep{Thorngren2018}. From the results of \citet{Miller2009}, an additional constant heat source of $10^{29}$ erg s$^{-1}$ could explain the inflated radius of WASP-103b. For this same additional heating, an inflated WASP-18b would have a radius of 1.3R$_{Jup}$ \citep{Miller2009}. This is marginally larger than the observed radius of WASP-18b by 2$\sigma$. Thus WASP-18b appears slightly less inflated than WASP-103b despite being around an almost identical star and at the same dayside temperature.
One inflation mechanism that can explain this difference is Ohmic dissipation as, for a fixed additional heating, it is less efficient and inflating the radii of a higher mass planet \citep{Huang2012}.

These two differences both highlight that there is a greater complexity behind the observed properties, such as the phase curve properties, in particular in the role that magnetic fields might play on the circulation or radius inflation. Importantly, we can use planets such as the UHJs as a test for inflation theories, as they occupy a parameter space where inflation models differ \citep{Sestovic2018, Thorngren2018}

\section{Conclusions}
\label{Sec:Conclusions}
We observed one full orbit phase curve of the ultra hot Jupiter WASP-18b. We find the peak signal from the dayside at an effective temperature of 2894$\pm$30 K and do not detect the nightside of the planet, placing an upper limit of 1430K at 3$\sigma$. We find a large day-night-contrast of >0.96 in luminosity and a small offset of the brightest point from the substellar point by 4.5$\pm$0.5 degrees. 

We compare the extracted spectroscopic phase curve with Global Circulation modelling and find that the data can be best reproduced by models with efficient drag. 
Models without additional drag sources fail to reproduce the day-night contrast seen in our data, hence we require an additional drag source to explain the observed day-night contrast. We also find that the behaviour of the phase curve of WASP-18b with metallicity is different from cooler planets, owing to the high temperature chemistry of TiO and water in the atmosphere of WASP-18b. In addition to this, we show that a metallicity enhancement or depletion in our models is not sufficient to match the observed day-night contrast without the presence of efficient drag.

We explore the origin of this efficient drag, and show that it could be due to Lorentz forces on ionized metals in the atmosphere from a magnetic field as weak as 10 Gauss. The effect of a magnetic field on the circulation may also explain the small brightness offset seen in our data, however our models do not explore the full dependence of the circulation on magnetic effects, which will require further studies.

Furthermore, we compare our results to the recently published phase curve of WASP-103b \citep{Kreidberg2018}. We find that the two planets are consistent with the expectation that more massive planets should be less inflated, and support the theory of Ohmic dissipation as an inflation mechanism. However, their different circulations point to a more complicated picture and suggest that other fundamental properties of these systems, such as their magnetic fields, may be different.

\subsection*{}
We thank N. Cowan for their thorough referee report that greatly improved the clarity of this work. We also thank D. Thorngren for providing tables of predicted planet radii used in this work, and C. Helling for many fruitful discussions.
J.M.D. acknowledges that the research leading to these results has received funding from the European Research Council (ERC) under the European Union's Horizon 2020 research and innovation programme (grant agreement no. 679633; Exo-Atmos). J.M.D acknowledges support by the Amsterdam Academic Alliance (AAA) Program.
Support for program GO-13467 was provided to the US-based researchers by NASA through a grant from the Space Telescope Science Institute, which is operated by the Association of Universities for Research in Astronomy, Inc., under NASA contract NAS 5-26555. J.L.B. acknowledges support from the David and Lucile Packard Foundation.


\newpage

\begin{table}[ht]
\begin{center}
\begin{tabular}{ | c | c | c |}
\hline
Fit parameter & Best fit value & Error \\
\hline
c$_1$ & 5.004e-04 & 5.060e-06  \\
c$_2$ & -1.041e-02 & 5.553e-03  \\
c$_3$ & 7.539e-05 & 2.856e-05  \\
c$_4$ & -7.993e-03 & 7.545e-03  \\
fp & 9.701e-04 & 1.133e-05  \\
E$_C$ & 1.847e-04 & 2.935e-05  \\
C$_{scan, f}$ & 6.791e+08 & 2.280e+04  \\
C$_{scan, r}$ & 6.783e+08 & 2.248e+04  \\
V$_1$ & -2.252e-03 & 4.915e-05  \\
V$_2$ & 1.467e-03 & 4.034e-05  \\
$\tau$ & 8.402e-03 & 1.510e-04  \\
$\Delta$t & 4.171e-03 & 5.049e-03  \\
R$_{orb, 1}$ & 1.943e-03 & 3.004e-05  \\
R$_{orb, 2}$ & 1.038e-03 & 2.658e-05  \\
R$_{orb, 3}$ & 1.275e-03 & 9.025e-06  \\
R$_{orb, 4}$ & 1.695e-03 & 2.651e-05  \\
\hline
\end{tabular}
\end{center}
\caption{Best fit values resulting from the white-light curve fit. Variables are defined below Equation~\ref{Eq:model}. Flat priors were placed on all the parameters within acceptable physical ranges. An additional prior was used, ensuring that the minimum of the phase curve (calculated with c$_{1-4}$ and fp) was non-negative. }
\label{Tab:Fits}
\end{table}

\begin{figure*}
\begin{center}
\includegraphics[scale=1]{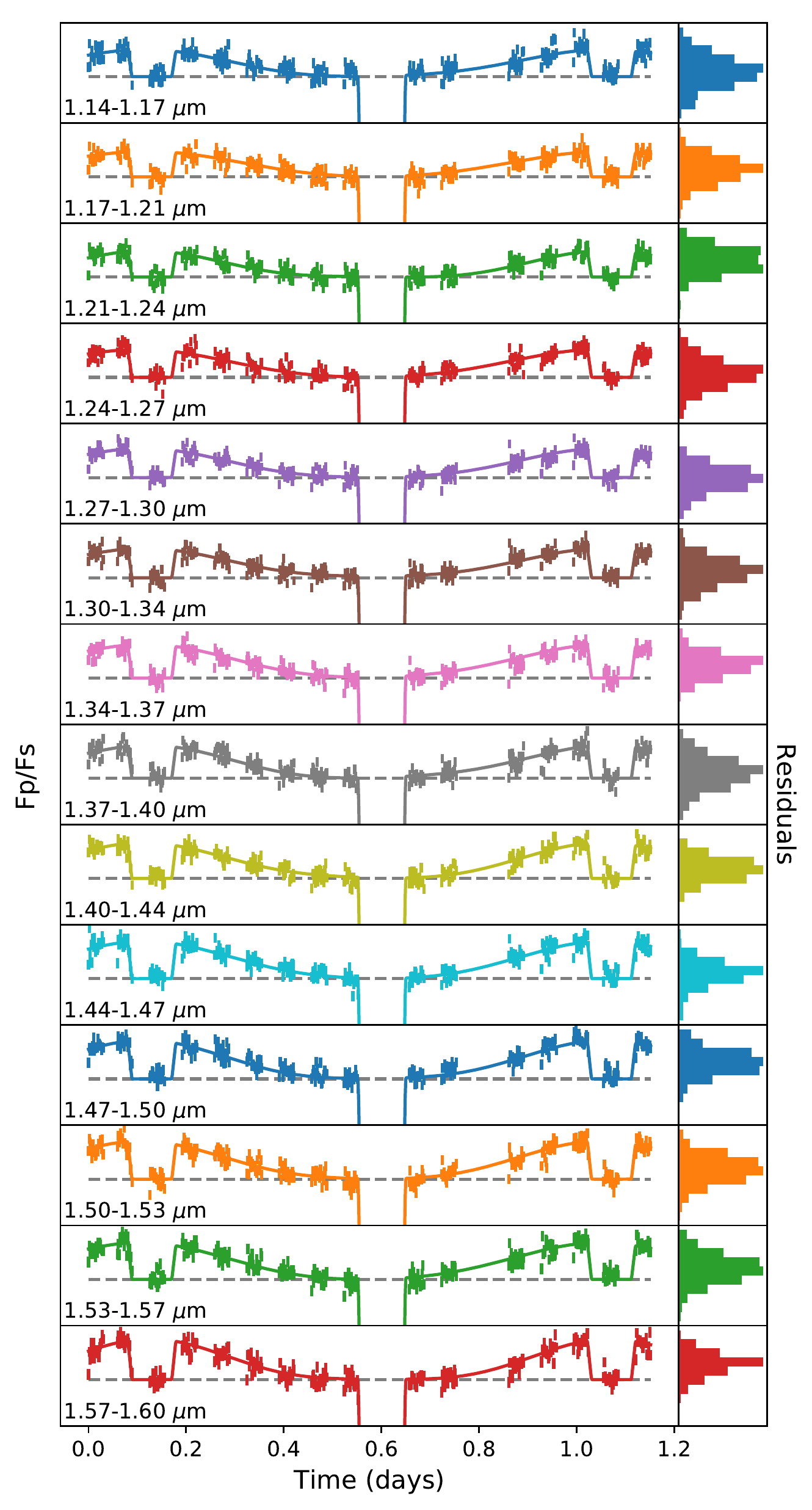}
\caption{Extracted spectroscopic phase curves shown with histograms of residuals. Solid curves indicate fits to the data.}
\label{fig:pcs}
\end{center}
\end{figure*}

\begin{figure*}
\begin{center}
\includegraphics[scale=0.2]{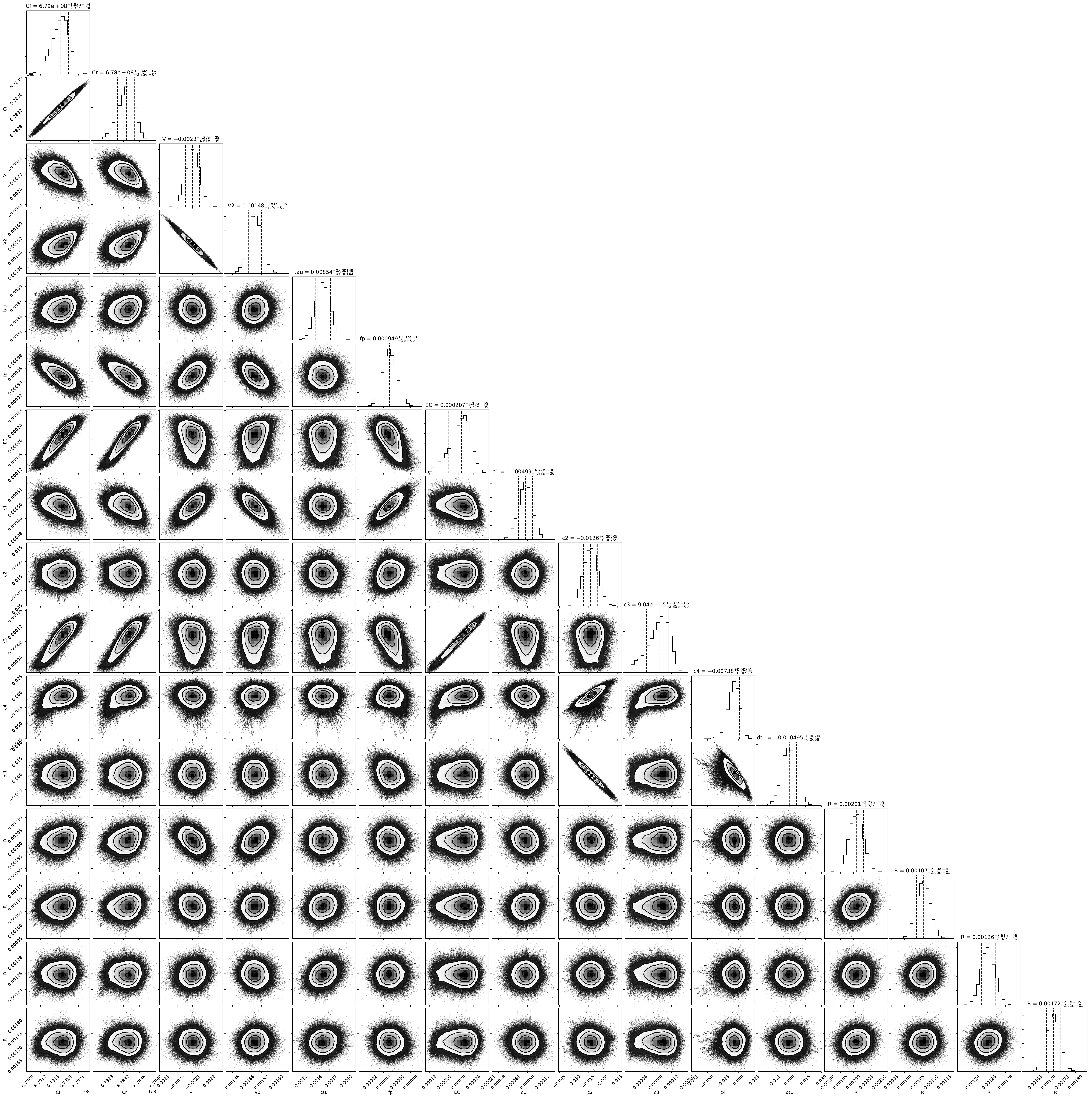}
\caption{Corner plot of posteriors for the white-light curve fit. Result of 10000 steps per 50 walkers. Generated using corner.py \citep{corner}. Histogram titles show means and $\pm1\sigma$ confidence intervals of the samples.}
\label{fig:corner}
\end{center}
\end{figure*}

\end{document}